\documentclass{PoS}

\title{Higgs Cosmology and Dark Matter}

\ShortTitle{Higgs Cosmology}

\author{\speaker{Ian G. Moss}\thanks{This work is supported by the Leverhulme Trust grant RPG-2016-233.}\\
        School of Mathematics, Statistics and Physics\\
       Newcastle University, Newcastle upon Tyne NE1 7RU, UK.\\
        E-mail: \email{ian.moss@ncl.ac.uk}}

\author{Ruth Gregory$^\dagger$
\thanks{Also supported by Perimeter Institute. Research at Perimeter 
Institute is supported by the Government of
Canada through Industry Canada and by the Province of Ontario through the
Ministry of Research and Innovation.}\\
Centre for Particle Theory, Durham University,
South Road, Durham, DH1 3LE, UK\\
Perimeter Institute, 31 Caroline Street North, Waterloo, 
ON, N2L 2Y5, Canada\\
        E-mail: \email{r.a.w.gregory@durham.ac.uk}}

\abstract{Higgs vacuum stability has important consequences for cosmology.
In particular, we argue that if the Higgs vacuum is metastable, then the dark
matter cannot contain a single black hole of mass less than $10^{15}{\rm g}$
in our entire past light cone. In addition to being the destroyer microscopic black holes,
it may be possible that Higgs vacuum decay is a source of
primordial black holes during inflation.}

\FullConference{2nd World Summit: Exploring the Dark Side of the Universe\\
		25-29 June, 2018\\
		University of Antilles, Pointe-à-Pitre, Guadeloupe, France}

\begin{document}

\section{Introduction}

Before the Higgs boson was discovered back in 2012, there were already
lower bounds on the Higgs mass set by requiring stability of the Higgs vacuum
against quantum corrections from the standard model fermions \cite{Degrassi:2012ry}.
As it turned out, the Higgs mass came out remarkably close to the lower bound,
making it likely that the Higgs vacuum of the standard model is a long-lived metastable state
\cite{Blum:2015rpa}. 
This situation has interesting cosmological consequences that we will address here.

At the very least, Higgs stability gives us limits on particle physics over a very wide
range of energy scales. Fermions and bosons work in opposite ways in stabilising
or destabilising the Higgs vacuum. The top quark mass is the most important uncertain
parameter in determining Higgs stability in the Standard Model. Physics beyond
the standard model must be constrained by Higgs stability.

In early universe cosmology, Higgs instability is suppressed at high temperatures so attention has focussed 
around the inflationary era when the universe was cooler \cite{Espinosa:2007qp,EliasMiro:2011aa}. 
During inflation, the gravitational couplings
of the Higgs boson play a crucial role. Higgs stability sets limits on the Higgs-gravity interactions,
but it also gives rise to the possibility of first order phase transitions. These produce
large density inhomogeneities, which in turn could produce relic particles and hence the connection
with dark matter in the title of this talk.

\section{Vacuum stability}

The basic idea of vacuum destabilisation can be illustrated very simply by looking at the
ground state energy of the standard model fields. Each momentum mode acts as a harmonic oscillator
with frequency $\omega$, but the ground state energy of the fermions is negative. The total
vacuum energy is
\begin{equation}
\Delta V=\sum_b\frac12\hbar\omega_b-\sum_f\frac12\hbar\omega_f.
\end{equation}
The frequencies depend on the effective particle mass, but this depends in turn on the
magnitude of the Higgs field $\phi$ and the mass $M$ in the Higgs vacuum,
\begin{equation}
\omega=(k^2+M^2\phi^2/v^2)^{1/2},
\end{equation}
where $v=246{\rm GeV}$. The largest effect comes from the most massive field,
which is the top quark. 

In practice, the sums are infinite and have to be regularised. When added to
the classical Higgs potential, the combined potential at ultra-high energy has a quartic form
with a slowly running self-coupling $\lambda(\phi)$,
\begin{equation}
V(\phi)=\frac14\lambda(\phi)\phi^4.
\end{equation}
Depending on the top quark mass, the potential becomes negative at some scale
$\Lambda$ which lies somewhere in the range $\Lambda>10^{10}{\rm GeV}$.
The true minimum of the potential lies beyond the Planck scale, but this
is beyond the range of validity of the effective potential.

\begin{figure}
\centering
\includegraphics[width=0.45\linewidth]{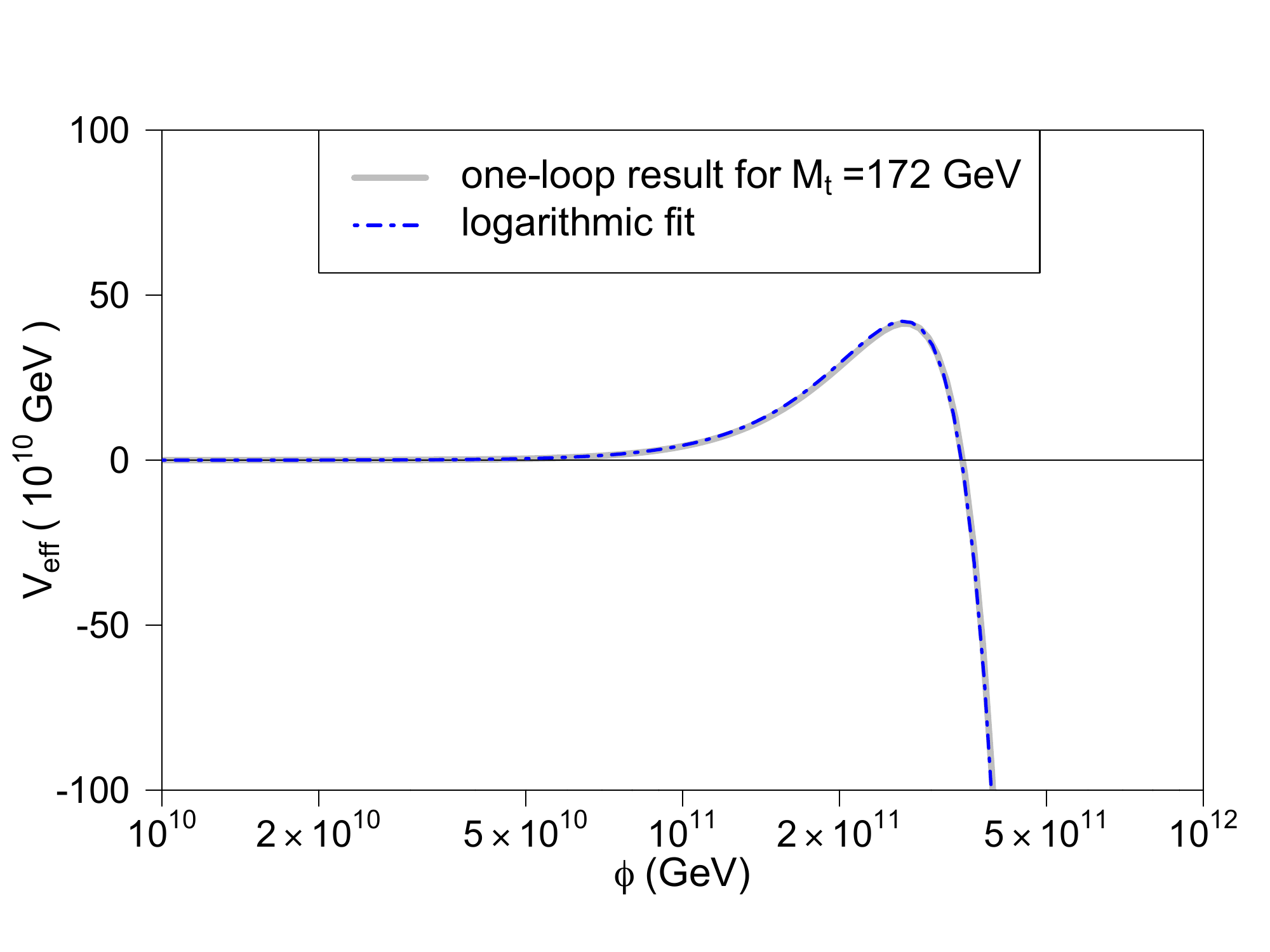}
\caption{The Higgs potential barrier from a two-loop calculation
with a top quark mass $172{\rm GeV}$. The Higgs potential vanishes 
at an intermediate scale $\Lambda$, in this
case $3\times 10^{11}{\rm GeV}$.} \label{fig:potential}
\end{figure}

\section{Present day vacuum stability}

Vacuum decay from the metastable, or false vacuum, state  is analogous 
to a first order phase transition that takes place through the nucleation of bubbles. 
The classic results of Callan and Coleman \cite{coleman1977,callan1977} 
tell us how to calculate the nucleation probability using 
imaginary time `bounce' solutions $\phi_b$.
The nucleation rate is given by
\begin{equation}
\Gamma=Ae^{-B},
\end{equation}
where $B$ is the change in the action between the bounce and the false vacuum. 
The pre-factor $A$ is a numerical factor of order $\Lambda^{-4}$
which we will ignore.
In these calculations, we include the gravitational back reaction and so there are terms
in the action for the Higgs field and for gravity.

\begin{figure}
\centering
\includegraphics[width=0.45\linewidth]{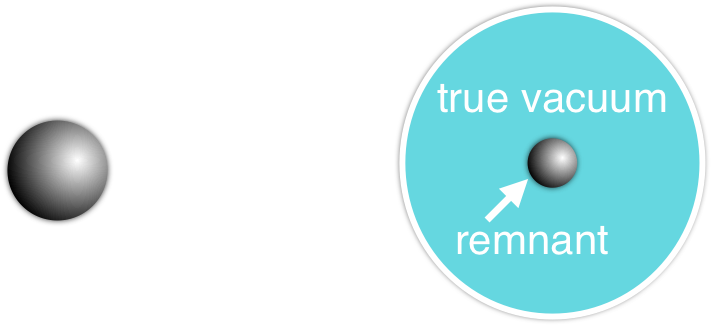}
\caption{Black holes act as nucleation seeds of vacuum decay  
} \label{fig:seed}
\end{figure}

All first order phase transitions, such as he condensation of water vapour,
are exponentially suppressed and the lifetimes of the metastable states are
typically huge. In practice, bubbles nucleate much faster around nucleation 
seeds such as grains of dust or defects in a containment vessel. The same is true
for Higgs vacuum decay, and in this case the nucleation seed may be a microscopic
primordial black hole as in Fig. \ref{fig:seed},

The nucleation rate can be calculated as before \cite{BGM1,BGM3,Gregory:2016xix}, 
but now $B$ is the difference in
action between the bounce solution and the black hole seed. Both the bounce and the
black hole seed should have the same mass as measured by a distant observer, and
consequently the mass of the black hole inside the bubble shrinks. The change
in action is given by a remarkably simple result,
\begin{equation}
B={1\over 4G} {\cal A}_{\rm seed}-{1\over 4G} {\cal A}_{\rm remnant},
\end{equation}
where ${\cal A}=16\pi M^2$ is the area of the event horizon for a black hole of mass $M$.
We know, however, that the area of the black hole and the entropy $S$ are related,
and hence the nucleation rate is given by the thermodynamic formula
\begin{equation}
\Gamma\propto e^{\Delta S}
\end{equation}
This is a beautiful example which demonstrates that the black hole entropy
formula applies to fluctuations as well as to thermodynamical equilibrium.

The relationship between the mass of the seed black hole and the remnant black hole
has to be evaluated numerically by solving the gravity-Higgs equations for the bounce
solutions. The vacuum decay rate is in competition with the another process, the
decay of the hole due to hawking evaporation. The numerical results show that, for a
metastable Higgs vacuum, the decay rate always exceeds the evaporation rate at small
black hole masses in the range $10^5-10^9$ times the Planck mass. 
Primordial black holes with mass greater than this evaporate and their mass
decreases. At some point, the vacuum decay rate exceeds the
evaporation rate, and the black hole seeds vacuum decay. This process happens whenever
a black hole is produced in the early universe and is sufficiently small that it evaporates before
the present epoch, which means a mass less than around $10^{15}{\rm g}$.
This has a consequence for dark matter. {\em If the Higgs is metastable, then the dark
matter cannot contain a single black hole of mass less than $10^{15}{\rm g}$
in our entire past light cone.}

\section{Early universe vacuum stability}

Higgs vacuum stability also has interesting implications for the early universe,
particularly during a period of early universe inflation. Consider a physical
theory that combines the Standard Model with General Relativity and an
inflaton field. This can be applied consistently as an effective theory
at energies below the Planck mass. We find that:

\begin{enumerate}
\item de Sitter vacuum fluctuations can destabilise the Higgs;
\item the curvature coupling $\xi$ plays a decisive role in vacuum stability;
\item the Higgs vacuum may decay during inflation, but recover during reheating.
\end{enumerate}

The scale of Higgs vacuum fluctuations during inflation is set by the Hubble expansion rate
during inflation $H$. In typical inflationary models, the Hubble parameter 
$H\approx 1.0\times10^{15}\epsilon_V\,{\rm GeV}$, where $\epsilon_V$ is a slow roll 
parameter. The fluctuations can easily be larger than the width of the
barrier in the Higgs potential, and there are limits on the Higgs potential,
or on inflationary scenarios, arising from stability of the Higgs vacuum.

The actual situation is complicated by the possibility of interactions between the
Higgs field and gravity or between the Higgs field and the inflaton $\Phi$. The
actual Lagrangian density is of the form
\begin{equation}
{\cal L}_{\rm eff}=-\frac14\lambda(\phi)\phi^4-\frac12\xi(\phi)R\phi^2-\frac12g(\phi)\phi^2\Phi^2+\dots
\end{equation}
Notice that the couplings are themselves dependent on the Higgs field because
of quantum corrections. They are not protected by any known symmetry and therefore
they cannot be zero for all values of the Higgs field. 

\begin{figure}
\centering
\includegraphics[width=0.45\linewidth]{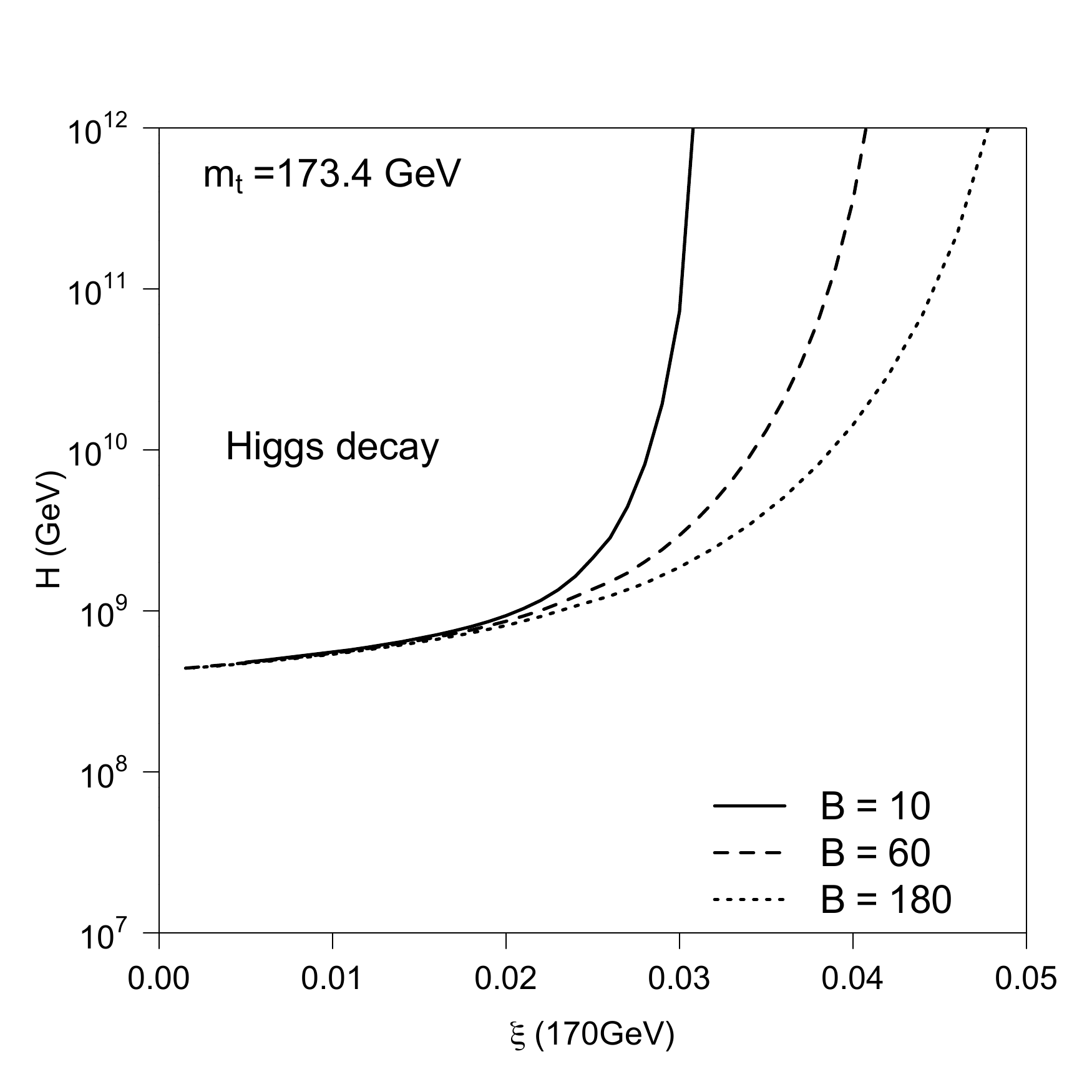}
\caption{The Higgs stability region during inflation (from  \cite{Bounakis:2017fkv} ). 
The axes are the Hubble parameter
during inflation and the curvature coupling at low energy. The
inflaton-Higgs coupling has been set to zero.
A top quark mass $173.4{\rm GeV}$ has been used.
The lines show different values of the tunnelling exponent. Higgs decay in any of the $e^{180}$ 
inflationary Hubble regions would spread throughout the entire universe.
} \label{fig:limits}
\end{figure}

The curvature coupling term  is overwhelmed by thermal corrections to the potential for most 
of the duration of the hot big bang, and the term is vanishingly small today. It plays
a crucial role in Higgs stability during inflation, as shown in Fig. \ref{fig:limits}. Positive values of $\xi$
increase the height of the potential barrier and stabilise the Higgs field, whilst
negative values destabilise the Higgs field \cite{Espinosa:2007qp}.

Fig. \ref{fig:limits} includes the effects of running in both the Higgs self-coupling and the curvature
coupling. There is a technical point in the calculation that is of some importance. First, we note
that the metric and Higgs potential can be redefined to remove the curvature coupling.
In so doing, we mix up the Higgs mass with the couplings, but we should not change
the physics. The quantum theory therefore has to be done in a way that is covariant under
such field redefinitions. This is a non-trivial exercise \cite{Bounakis:2017fkv}. However, there are combinations of
parameters that can be easily calculated, for example if the Higgs mass square is $\mu^2$,
then the relevant combination is $\mu^2+\xi R$.

Thus, depending on the parameters in the potential and the Hubble constant  during inflation, the
Higgs vacuum may tunnel through the potential barrier in Fig \ref{fig:potential}. Once through, 
however, the Higgs
takes some time to role down the potential so that inflation can end during this Higgs role phase,
and thermal effects from the reheating universe could re-stabilise the Higgs field. The Higgs could
therefore undergo a first or second order phase transition during inflation. In
 the second order case, it has been suggested that perturbative Higgs fluctuations during the 
 roll-down phase could seed primordial black holes as possible dark matter candidates
 \cite{Espinosa:2017sgp}. If the decay is first order, bubbles may collide to form black holes 
 by the mechanism first proposed in \cite{PhysRevD.26.2681,Kodama:1982sf}.
 
 \section{Acknowledgements}
 
We would like to thank the organisers and sponsors of the 2nd World Summit: Exploring the Dark 
Side of the Universe. We are also grateful for the collaboration of Ben Withers and Philipp Burda
on black hole nucleation, and to Marios Bounakis on the gravitational Higgs
corrections.

\end{document}